\let\FONTS=C
\catcode`\{=1 
\catcode`\}=2 
\catcode`\#=6 
\let\x=\input
\def\y#1 {\let\input\x \let\x=\undefined \input lfonts.new}
\def\input#1 {\let\input=\y \let\y=\undefinded}
\x lplain

\input clmult01.mac
\makeatletter
\input amsfont.sty
\input cropmark.sty
\nopagecropped
\makeatother

\documentstyle[fleqn,psfig]{clmult01}

\begin{document}
\title{Modelling Quasicrystal Growth}
\author{Uwe Grimm\inst{1} \and Dieter Joseph\inst{2}}
\institute{Institut f\"{u}r Physik, 
           Technische Universit\"{a}t Chemnitz,\\
           D-09107 Chemnitz, Germany
           \and
           Laboratory of Atomic and Solid State Physics,
           Cornell University, \\
           Ithaca, NY 14853-2501, U.S.A.}
\authorrunning{Uwe Grimm and Dieter Joseph}
\titlerunning{Growth Models}
\maketitle

\noindent {\em Dedicated to the memory of Marko V.\ Jari\'{c}}

\section{Introduction}

How do quasicrystals grow? This has been (and still is) one of the
fundamental puzzles in the theory of quasicrystals, and despite the
effort employed to provide an answer to this fascinating question,
there is yet much to be understood. In this review, we want to show
the problems one has to face, and present an overview on the ideas,
the models, the methods, and the results that have been discussed in
the literature to date.

But, after all, one may ask, what is so different about growing
quasicrystals in comparison to the growth of (ordinary) crystals?
Certainly, taking a thermodynamic point of view, one would expect that
the basic theory of nucleation and crystal growth should be the same
as for ordinary crystals; see e.g.\ \cite{Markov} for an introductory
textbook on this subject, and \cite{HolMo} for experimental results on
undercooled melts of quasicrystalline alloys.  In fact, the answer to
this question lies in the structure of quasicrystals, and their
relationship to ordered crystals, on the one hand, and disordered
amorphous materials, on the other.  An ideal crystal is constructed as
a periodic arrangement of a single building block, the unit cell,
forming a regular lattice. Thus, in order to understand why the atoms
arrange in a crystalline fashion, it is sufficient to consider a small
portion of the crystal containing a few of these unit cells. In
amorphous systems, on the other hand, one only finds short-ranged
local order, and one can quite easily imagine how such a structure can
emerge in a process of aggregation.

For {\em ideal}\/ quasicrystals (where ``ideal'' means perfect in the
mathematical sense), however, the situation is completely
different. Being aperiodic, they possess a long-ranged quasiperiodic
order of the atomic positions. This quasiperiodic order, if it is
perfect, is very restrictive, and one should not think of the
quasiperiodic structure as being less ordered than the periodic
structure that underlies crystals. In fact, structure models of
quasicrystals are typically built on quasiperiodic tilings of space,
where the tiles are decorated by the atomic positions. Such tilings
can be obtained as sections through higher-dimensional periodic
lattices, hence the degree of order in these periodic and
quasiperiodic structures is comparable. But the lack of periodicity
makes it much more difficult to build such a structure from its
constituents (for example, to construct a quasiperiodic tiling by
arranging the tiles like a puzzle) without violating quasiperiodicity,
compare \cite{Penrose}. We cordially invite the reader to give it a
trial, see \cite{GS} for details.

At first, the tile-by-tile growth of a quasicrystal may even seem to
be an impossible task. However, for certain tilings, there exist
collections of {\em local}\/ conditions, so-called matching rules,
that enforce quasiperiodicity. Nevertheless, these matching rules do
not really solve the problem: they only state that if you managed to
tile the space without violating the matching rules, you will indeed
have constructed a quasiperiodic tiling. But they do not guarantee
that a given finite ``legal'' patch (i.e., one that does not violate
the rules) {\em can}\/ in fact be extended to an infinite tiling of
space.  So, in this sense, matching rules are not sufficient to
provide a simple algorithm for the construction of ideal quasiperiodic
tilings.

With this in mind, one inevitably arrives at the question of how
Nature solves this puzzle to grow real quasicrystals. Apparently, this
seems to require non-local interaction among the constituents which we
want to exclude on physical grounds.  Of course, there is a rather
easy way out of this dilemma -- namely that in physical reality one
{\em never}\/ encounters real quasicrystals that are absolutely
perfect in the mathematical sense. But the same is true for ideal
crystals, and one can nowadays grow real quasicrystals that show at
least comparable degrees of perfection to that of real ordinary
crystals -- judged, for instance, by the sharpness of their
experimentally observed diffraction patterns. These high-quality
quasicrystals are commonly called ``perfect quasicrystals'' regardless
whether they resemble an ideal quasiperiodic tiling or a random tiling
(see below).

Nevertheless, it turns out that the idea of building defects into the
ideal quasiperiodic tilings is rather fruitful, being closely linked
to the question of thermodynamic stability of quasicrystals. In
principle, there are two possible scenarios: quasicrystals may either
be true ground states of the many-particle Hamiltonian, and hence {\em
energetically}\/ stabilized (which means that they are
thermodynamically stable at arbitrarily low temperatures), or they are
high-temperature phases which are stabilized by {\em entropic}\/
contributions to the free energy. In the latter case, which is
tentatively favoured by experimental results (see
\cite{deBo95,Boudard,JBR97}), the question of disorder in
quasicrystals is crucial since disorder provides entropy. Examples
are, for instance, defects and vacancies (as also frequently observed
in crystals), configurational disorder (which can be described by
randomizing the ideal quasiperiodic tiling, resulting in random tiling
models), or chemical disorder (which emerges by interchanging atomic
positions between different atomic species).  Arguably, all these
sources of entropy appear in quasicrystals, and there is experimental
evidence that the description of quasicrystals as random tilings may
be more appropriate than those based on ideal quasiperiodic tilings,
see also \cite{Nissen}. Most of the models in the literature therefore
do not describe the growth of ideal structures, but rather try to grow
tilings incorporating defects, or structures resembling random
tilings.

This review is organized as follows. First, we concentrate on ideal
qua\-si\-per\-i\-od\-ic tilings, discussing the possibility of growing
ideal Penrose tilings as an example. This will also clarify the
question of non-locality in ideal quasiperiodic growth.  After that,
we will briefly address the basic problem of random growth models.
Subsequently, a number of semi-realistic growth models is discussed
which try to mimic the growth of defective quasiperiodic structures in
different ways. We distinguish altogether four different classes of
growth models, but we would like to stress that our somewhat arbitrary
distribution of models into these groups serves mainly presentational
purposes and is not meant as a real classification.

The first group consists of what we refer to as ``atomistic'' growth
models, where single ``atoms'' are added to the growing cluster
according to certain conditions. The second class contains
(algorithmically motivated) ``cluster-based'' models, where the
aggregation is driven by the requirement that certain preferred
cluster configurations have to be completed. After this, we address
more physically motivated cluster-based models, so-called
``orientational glasses'', which are random packings of (symmetric)
building blocks preserving relative orientations, and hence show
orientational order. The last group of growth models in our list is
based on the random tiling idea, starting from a fixed set of
tiles. Finally, we conclude and give an outlook on future
developments.

\section{Growing Ideal Penrose Tilings}

There has been a hot debate on the possibility of purely local growth
al\-go\-rithms for ideal quasiperiodic tilings (their inherent
non-locality was first pointed out by V.~Elser \cite{Elser85}),
particularly after Onoda et al.\ \cite{OSDS88} found an algorithm for
the Penrose tiling and claimed that it was local (see the comment by
Jari\'{c} and Ronchetti \cite{JarRon}, the reply of Onoda et al.\
\cite{OSDS89}, and the beautiful article by Penrose
\cite{Penrose}). In what follows, we shall discuss this algorithm,
occasionally referred to as the ``OSDS rules'', in some detail,
compare also the rather nice presentation by Socolar \cite{Socolar},
and review articles by Steinhardt \cite{Steinhardt} and DiVincenzo
\cite{DiVincenzo}.

\begin{figure}
\sidecaption
\psfig{file=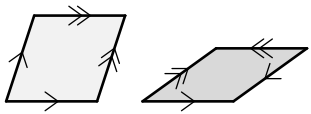,width=42mm}
\caption{The Penrose rhombs with arrow decorations.}
\label{fig1}
\end{figure}

As one may have guessed from our introductory comments, the algorithm
is based on the matching rules for the Penrose tiling. These {\em
perfect}\/ matching rules (see \cite{Baake}) are most easily described
in terms of two different types of arrows (``single'' and ``double''
arrows) assigned to the edges of the two rhombs (``fat'' and ``thin''
rhombs) as shown in Fig.~\ref{fig1}.  Starting with some initial
patch, one successively adds single tiles to the surface obeying the
matching rules. But, as mentioned above, this is not sufficient -- in
general, sooner or later, one arrives at the situation where a certain
surface vertex cannot be completed without violating the matching
rules.

\begin{figure}
\sidecaption
\mbox{\psfig{file=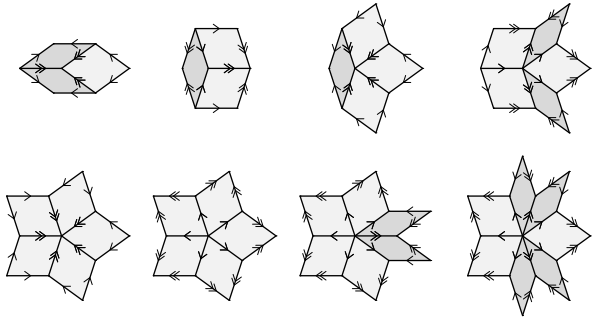,width=80mm}}
\caption{The eight vertex configurations of the Penrose tiling.}
\label{fig2}
\end{figure}

Thus, one has to be more careful about adding tiles, and in the OSDS
rules, going back to a suggestion of Gardner \cite{Gardner}, this is
achieved by distinguishing two kinds of tile additions: so-called
``forced'' and ``unforced'' tiles. As the name suggests, the first
kind of addition is one where there is only a {\em single}\/
possibility of adding a tile (in general, there will be two) at this
place without creating a vertex configuration that is not allowed in
an ideal Penrose tiling (there are only eight such configurations, see
Fig.~\ref{fig2}).

\begin{figure}
\mbox{
\psfig{file=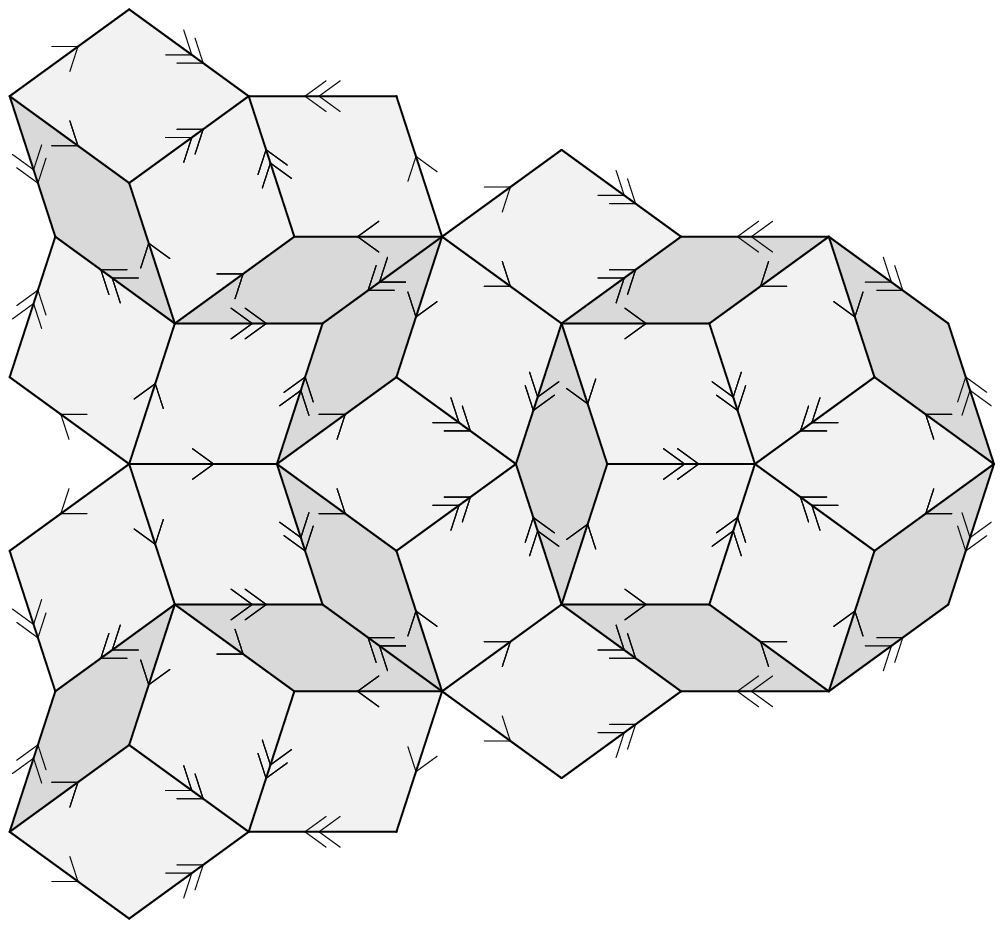,height=43.4mm}\hspace*{10mm}
\psfig{file=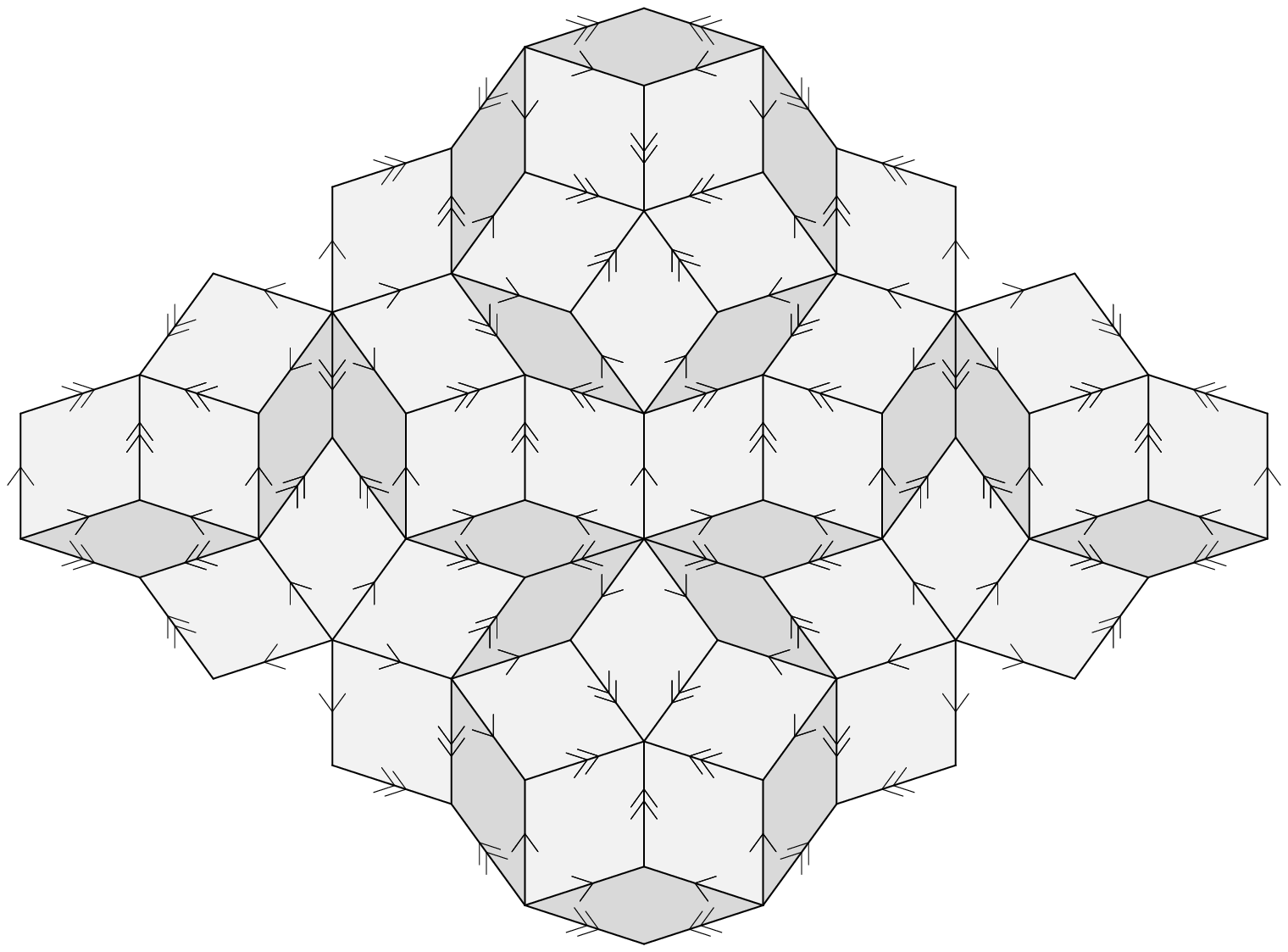,height=43.4mm}}
\caption{Two examples of a dead surface.}
\label{fig3}
\end{figure}

Growth according to the OSDS rules now proceeds in two sequential
steps. In the first step, one only adds forced tiles at randomly
chosen positions, until the entire surface consists of unforced
vertices, i.e., incomplete vertices at which no additions are
forced. Such surfaces are termed ``dead surfaces'', two small examples
are shown in Fig.~\ref{fig3}. There exist dead surfaces of arbitrary
size which can be catalogued exhaustively, see e.g.\ \cite{Socolar}.

\begin{figure}
\sidecaption
\mbox{\psfig{file=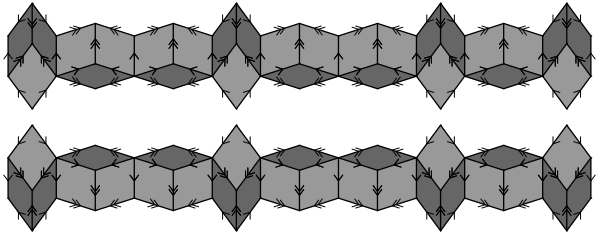,width=80mm}}
\caption{Two worm segments related by a flip.}
\label{fig4}
\end{figure}

\begin{figure}
\mbox{
\psfig{file=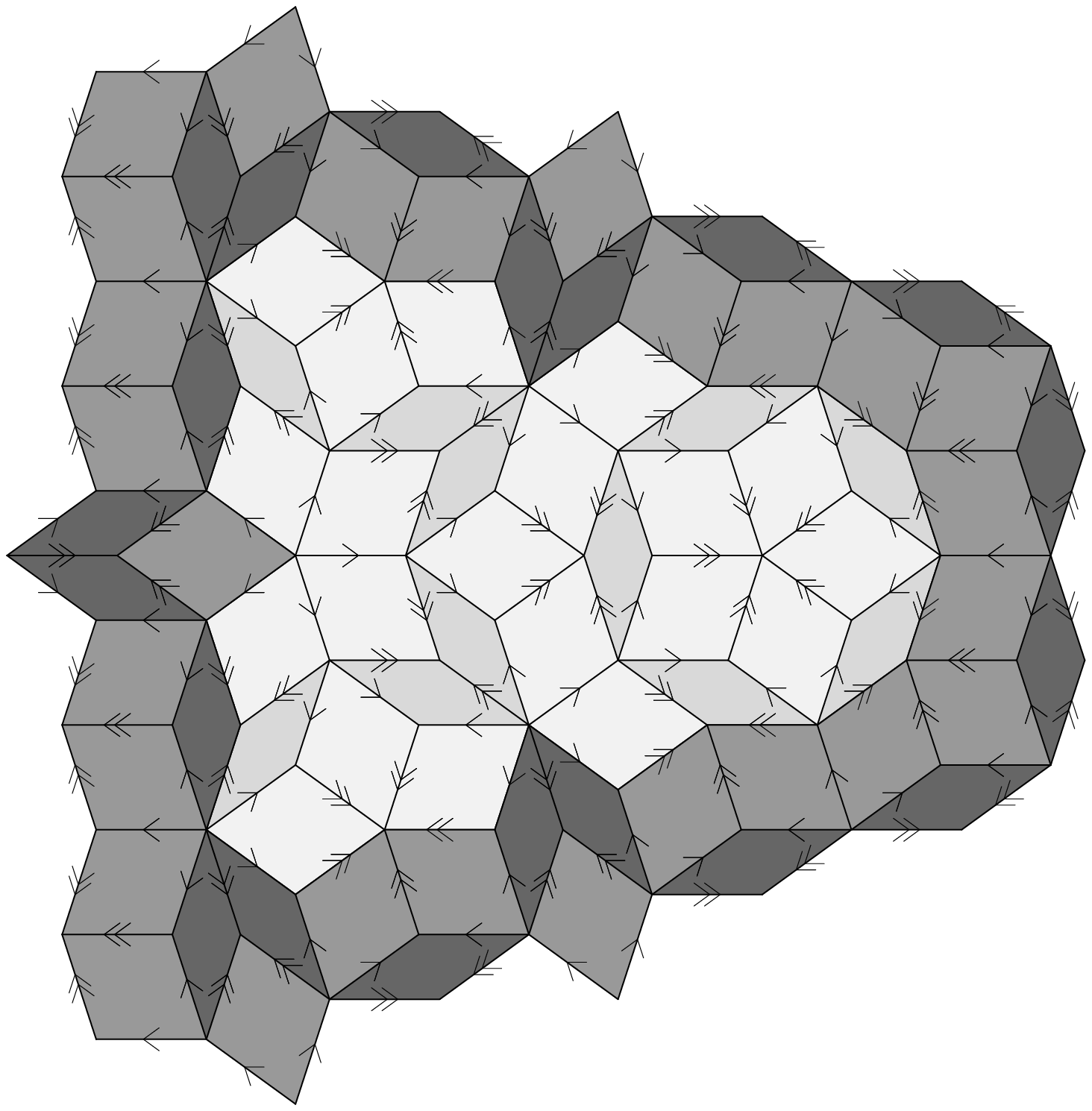,height=57mm}\hspace*{4mm}
\psfig{file=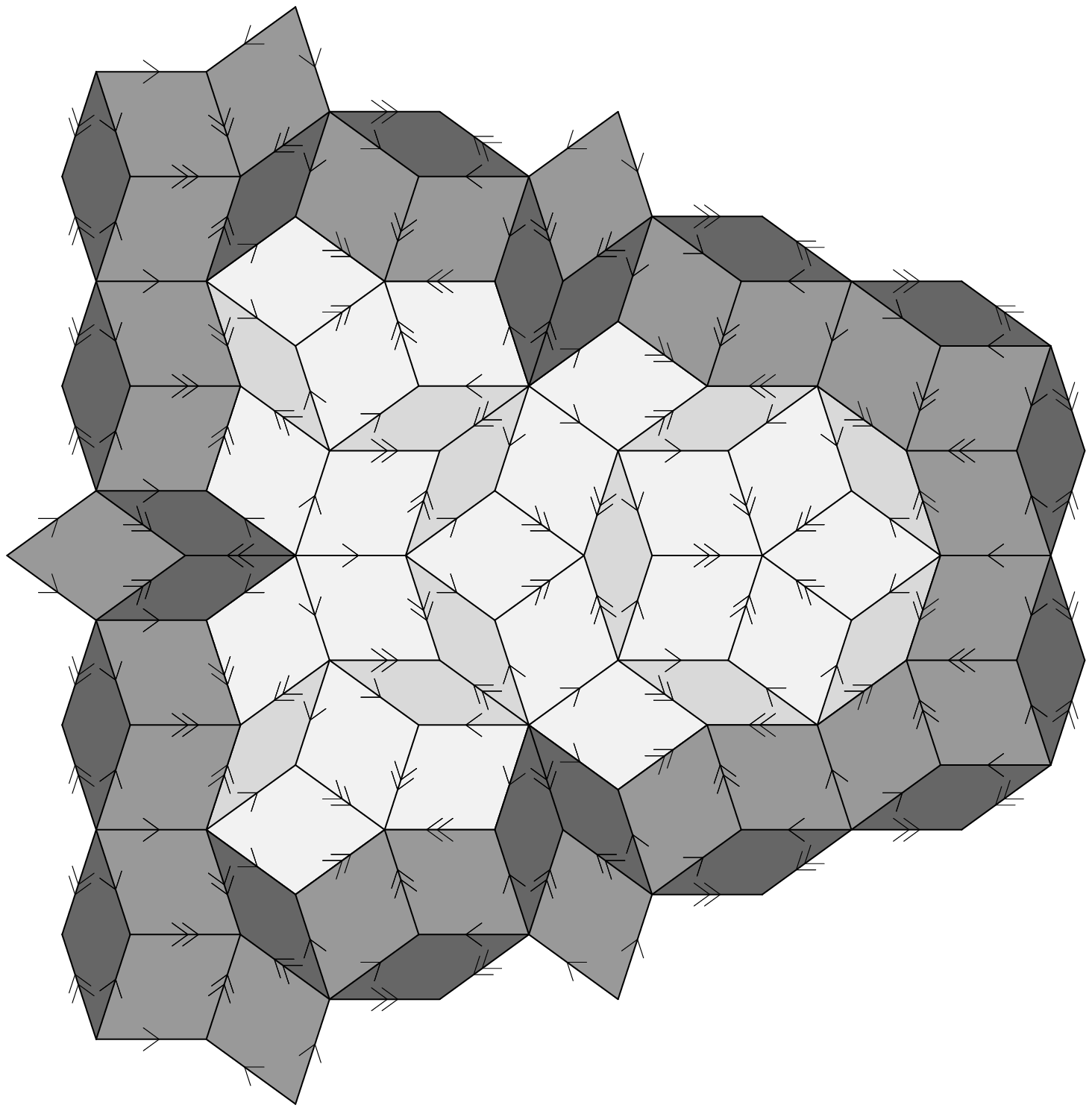,height=57mm}}
\caption{Two possible extensions of a dead surface by worm segments.}
\label{fig5}
\end{figure}

Along the flat parts of a dead surface, one may add segments of
so-called ``worms'' which consist of a linear chain of the two
hexagons containing three rhombs (Fig.~\ref{fig4}) arranged in a
Fibonacci sequence (see \cite{Penrose} for details).  The complete worm
can be ``flipped'' without affecting the form or arrow decoration of
its surface, hence it fits on the dead surface in both orientations,
see Fig.~\ref{fig5} for an example. 

While this shows the freedom one has to continue the growth process,
simply adding worm segments of arbitrary orientations to a dead
surface can lead to inconsistencies (there are ``dangerous faces'' of
the dead surface) which may only show up after the addition of many
more tiles, see \cite{Penrose}. Thus, in order to avoid this, one has
to formulate a more restrictive rule for adding the next tile, and the
second OSDS rule states that one should add a ``fat'' rhomb at either
side of an $108^{\circ}$ corner of the dead surface, consistent with
the vertex set of Fig.~\ref{fig2}. This gives one possible rule that
ensures that one never encounters inconsistencies (see \cite{Socolar}
for a more thorough discussion). Note that any dead surface has at
least two $108^{\circ}$ corners, except for certain degenerate cases
where two such corners merge to form a $36^{\circ}$ corner. 

How does non-locality in growth manifest itself?  The crux is hidden
in the two-step process: in order to avoid mistakes, one first has to
complete a dead surface before making an unforced choice.  However,
this implies that one has to inspect the {\em entire}\/ surface of the
patch. In other words, vertices at arbitrarily large distance have to
be visited as the patch grows.  In this sense, the OSDS growth
algorithm is clearly non-local, as pointed out by Jari\'{c} and
Ronchetti \cite{JarRon}, by Penrose \cite{Penrose}, and by Olami
\cite{Olami}.  Beyond this, it is generally not possible to judge
locally whether an unforced choice will yield inconsistencies or not,
see \cite{Penrose}.

However, it is indeed possible to derive a fully local algorithm if
one is willing to pay the price of defects, for instance by allowing
unforced choices with a very small probability. In this way, tilings
with an arbitrarily low defect rate (depending on the chosen
probability) can be grown.  However, as observed in \cite{RonJar}, the
growth of the tiling shows a rather strange time-dependence: while
forced tiles are added very quickly, growth becomes completely stalled
when a dead surface is approached. Only after an unforced tile is
chosen does rapid growth commence again.  A number of more general
growth rules, allowing defects of various types, was investigated by
van Ophuysen et al.\ \cite{OWD}.

Let us close the discussion of ideal tiling growth with two remarks.
The first is a warning concerning dead surfaces. It is by no means the
case that nothing outside a dead surface is determined by the ideal
structure -- in general, large parts of the tiling beyond the worm
segment are already fixed. The second remark concerns special kinds of
defects, so-called ``decapods'', which are decagons that cannot be
filled without violating the matching rules. Starting from such a
defect as an initial patch for the growth, one never encounters a dead
surface, and forced growth of an (otherwise) ideal structure continues
indefinitely, see \cite{Socolar}.

\section{Growth in the Random Scenario}

Beyond the essentially deterministic growth of ideal tilings,
non-locality usually poses no problem (for exceptions refer
to Sect.~7). Instead of restrictive matching rules, one has to fulfill
simple geometric constraints which guarantee symmetry, face-to-face
conditions, etc. New atoms, clusters, or tiles can be attached to the
growing surface using purely local information depending only on a
limited number of neighbouring objects. The choices in the growth
process correspond to the configurational entropy of the system.

Nevertheless, this scenario also has its drawbacks as was pointed out
by Sekimoto \cite{Sekimoto} and by Kalugin \cite{Kalugin}. The surface
aggregation suffers from a memory effect generating a considerable
rise in the background strain of the ``internal'' (also referred to as
``orthogonal'', ``perpendicular'', or ``phason'') coordinates, a
so-called phason strain. This terminology refers to the embedding into
higher-dimensional space, see \cite{Henley91,Baake,Trebin} and the
literature quoted therein. In fact, the corresponding phason
fluctuations become size-dependent for three-dimensional (3D)
quasicrystals. In practice, this means that a grown system shows
strong {\em systematic}\/ deviations from both ideal structure types
(ideal quasiperiodic tilings and equilibrium random tilings) rather
than the desired random fluctuations in the phason field alone.  The
behaviour can be modelled by a classical Ginzburg-Landau-Langevin
equation for the coarse-grained phason field $\Phi(r,z)$, $z$ being
the growth direction, $r$ the surface coordinates,
\begin{equation}
\frac{\partial{\Phi(r,z)}}{\partial{z}} =
{\cal D} \frac{\partial^2{\Phi(r,z)}}{\partial{r}^2} + \rho(r,z) \; ,
\end{equation}
where $\rho(r,z)$ represents a $\delta$-correlated noise term and
${\cal D}$ is the pha\-son-dif\-fu\-sion constant. This diffusion
equation leads to an asymmetry in the fluctuations of the phason
coordinates and hence in the spectral density. The intensities of the
Fourier spectrum of a grown 3D quasicrystal thus show a power-law
decay, the fluctuations of the internal coordinates diverge
logarithmically. Altogether, one ends up with an effective
``$D\!-\!1$ model'', i.e., the system grown in $D$ physical
dimensions behaves like a $(D\!-\!1)$-dimensional equilibrium
system. For details, we refer to \cite{Sekimoto}, an overview is given
by Henley \cite{Henley91}.

Most of the growth models discussed below belong to this class.
Although divergent phason fluctuations have been reported for
relatively-poor-quality quasicrystals, this is clearly not the case
for very-high-quality quasicrystals. On the other hand, the quality of
even the best samples benefits from annealing procedures, which means
that internal rearrangements over time, after the quasicrystal is first
grown, tend to increase the structural order. A possible scenario to avoid 
diverging phason strains is discussed in Sect.~7.

\section{Atomistic Growth Models}

Envisaging physical growth, the simplest model description grows a
cluster atom by atom, for instance by a diffusion-limited aggregation
process. In view of the peculiar properties of quasiperiodic
structures, this approach appears to be problematic, because it is
difficult to find simple assumptions (for example on interaction
potentials) that will result in the growth of a quasicrystal rather
than a periodic or a disordered structure. Monte Carlo investigations
hint at the possibility of stable quasicrystalline order, see e.g.\
\cite{WidStrSwe} for a two-component system with Lennard--Jones pair
potentials.  In fact, the models discussed below solve the problem by
restricting the possible positions where new atoms may be added to the
growing cluster, in this way trying to enforce non-crystallographic
symmetries or, at least, preventing simple periodic arrangements. In
some cases, these assumptions may appear rather artificial and hence
unphysical. Clearly, as for all models based on random aggregation
processes, one cannot expect to grow ideal quasiperiodic systems, but
rather more or less disordered structures; and the degree of disorder
present in the grown samples gives a first test of the relevance of
the model.

The earliest model of this kind was inspired by the binary decoration
of Penrose rhombs with two sizes of atoms. Equilibrium states of this
system had been studied by Lan\c{c}on et al.\ \cite{LanBilCha} and by
Widom et al.\ \cite{WidStrSwe}.  Minchau et al.\ \cite{MinSzeVil} and
Szeto and Wang \cite{SzeWan} consider a very simple growth model where
atoms may stick to certain positions determined by the tile decoration
of \cite{LanBilCha}. After being added to the growing cluster atoms
remain frozen at their positions. The growth is forced to proceed
``layer by layer'', new positions being taken into consideration only
after a full surface layer has been completed. The possible sticking
places for atoms are ordered according to an energy that is a sum of
pairwise contributions from neighbouring atoms (``growth-priority
parameters''), the available sites being filled successively according
to this ordering. In this way, the only freedom occurs when several
such places are degenerate in energy, in which case a random choice is
made.  As one might expect from such a simplistic model, the outcome
is not very close to a realistic quasicrystalline structure. In most
cases, the growing cluster generates ``tears''.  However, as shown in
\cite{SzeWan}, for certain ranges of the energy parameters the
layer-by-layer growth results in dense structures without holes or
tears (Szeto and Wang \cite{SzeWan} refer to these as ``perfect
structures'', a term that we want to avoid since it is prone to
misinterpretation). However, these grown samples resemble multiply
twinned crystalline structures much more than quasicrystals, showing
huge local anisotropies.

The growth model by Olami \cite{Olami} follows a somewhat different
strategy.  It starts from a small cluster of points from an ideal
quasiperiodic set, which in this case was not the Penrose tiling, but
a simpler pentagonal tiling obtained by projection. In such a finite
patch, there will be straight lines in the pentagonal directions
(``pentagonal lines'') that pass through a number of points in the
set. As possible positions of new points only intersections of such
pentagonal lines are considered; and a point is added on a randomly
chosen intersection of pentagonal lines if, in fact, it falls on at
least four lines. In addition, there is a rule that avoids short
distances, discarding the point with the smaller ``line number'' if a
new point comes too close to an existing point in the set (in the case
that the maximum number, i.e.\ five, pentagonal lines pass through
both points, both are kept). These growth rules are fully local,
depending only on the environment of a chosen point up to a distance
that is at most $\tau^3$ times the minimum distance of two points
along a pentagonal line in the ideal structure ($\tau$ is the golden
mean). Provided one starts with a sufficiently large initial cluster
(of about 30--100 points), one can grow large patches with rather
small fluctuations in the phason coordinates, resulting in very narrow
diffraction peaks in the Fourier transform. Furthermore, the phason
fluctuations grow extremely slowly with increasing system size, and
the numerical data from five grown patches with up to $2\cdot 10^4$
points suggest that they may stay bounded. Finally, Olami \cite{Olami}
points out that, in principle, a similar growth procedure could also
be applied to the Penrose tiling. Due to the more complex shape of the
acceptance domain one expects that grown patches will have larger
disorder than in the simpler pentagonal case.

\begin{figure}
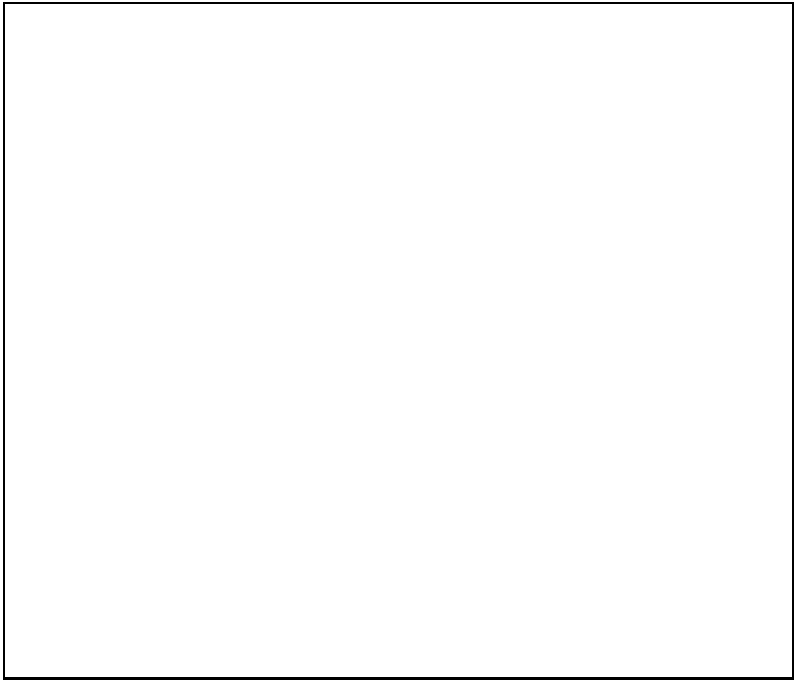

\mpicplace{105mm}{90mm}
\caption{View along a threefold axis of a DLO grown cluster of about
one million atoms (courtesy of V.~Dmitrienko).}
\label{fig6}
\end{figure}

As the last example is this category, we discuss the dodecahedral
local ordering (DLO) model of Dmitrienko and Astaf'ev \cite{DmiAst}
for the growth of icosahedral quasicrystals. Here, DLO means that all
closest neighbours of an atom are positioned at a subset of the
vertices of a regular dodecahedron. Such a local ordering may result
from a close packing of two kinds of atoms with different
sizes. Again, atoms once stuck to the cluster are frozen and cannot
move or be removed from the growing cluster. Starting from a seed of
about 3--50 atoms, trial positions at all vertices of regular
dodecahedra around all atoms are taken into account. The position for
a new atom is then selected by an energy that is a sum of
contributions for each atomic bond, depending only on atomic
distances. Dmitrienko and Astaf'ev \cite{DmiAst} chose their energy
parameters such that frustration between short- and medium-range
interatomic forces results, mimicking oscillating interatomic
potentials. In this way, the growth of crystalline structures is
suppressed.  Although the energy of a position can be obtained
locally, the whole surface must be inspected to find the position of
minimal energy. Fig.~\ref{fig6} shows an example of a large cluster
grown with these rules. The obtained samples are faceted, they have a
face-centered icosahedral structure, and their phasonic disorder
depends on the particular interatomic potentials.

\section{Algorithmically Motivated Cluster Growth Models}

The importance of clusters for the understanding of the structure and
the physical properties of quasicrystals was recognized immediately
after their discovery; but particularly over the past few years there
have been increasing research activities on a cluster-based
description of quasicrystals, see e.g.\ \cite{JaBoi,JeoSt94,DuJa}.
While, at first, this may seem to be in conflict with the tiling
picture, this is clearly not the case. On the one hand, one may
decorate tilings such that certain atomic clusters predominate the
local arrangements of atoms. On the other hand, at least some
quasiperiodic tilings, as for instance the Penrose tiling, can
alternatively be described in terms of a covering of space by a {\em
single}\/ cluster, see \cite{Duneau,Gummelt95,Gummelt96}. That is, in
contrast to the tiling picture, only a single tile is used, but
different kinds of overlaps may occur. For the example of the Penrose
tiling, it has been shown that a suitable restriction on the possible
overlaps of regular decagons is in fact equivalent to the perfect
matching rules.  The corresponding picture of a quasicrystal is
physically very attractive as one may imagine the quasicrystal as sort
of a densely-packed conglomerate of very stable clusters. Recently, it
was argued that a quasicrystalline structure may even result from a
{\em maximization}\/ of certain cluster configurations (see
\cite{GaJeo95,StJeo96,JeoSt97,Henley}), which may also give some new
insight into quasicrystal formation.  The growth models considered
here are of a different nature -- they are based on the addition of
single ``atoms'' guided by conditions that certain cluster
configurations have to be completed. In this sense, they are in fact
``atomistic'' models. However, the role of clusters as the fundamental
entities of the structure is strongly pronounced, which is the reason
why we want to consider them separately.

The first model we want to mention here, the so-called
decahedral-recursive (DR) model, was considered in a series of papers
by Romeu and coworkers \cite{Romeu,ArRoGo,RomAra}. In a sense, this
model bears some similarity to the DLO model \cite{DmiAst}, but it is
more closely tied to the cluster idea. The growth according to the DR
model proceeds in two steps. First, there is a ``decahedral stage'',
in which atoms are added to an initial seed (which the authors chose
to be magic number clusters of atoms, known to be particularly stable)
such that they complete irregular decahedra (pentagonal
bipyramids). This process is carried out with atoms of one size until,
due to geometric frustration, it becomes necessary to use a second
type of somewhat (about 5\%) smaller atoms, hence at least two atomic
species are required.  Note that frustrations necessarily appear due
to the different lengths of the edges and the outer radius of an
icosahedron (dodecahedron).  During this growth, certain atomic
positions (``O points'') on the surface occur that may act as new
nucleation centers. These are located at the center of (distorted)
icosahedra.  These O points have the property that, shifting the whole
cluster into them, a large number of coincidences are created. This is
exploited to enlarge the cluster further (discarding overlapping
atoms), until one arrives at particular O points (``nodes'') which
give rise to a larger percentage of coincidences than others. The
cluster obtained in this way is considered the ``basic cluster'' of
the structure, and further growth proceeds by shifting this basic
cluster into its nodes that maximize the fraction of coinciding atomic
positions. Thus, the grown structure may be thought of as a packing by
overlapping copies of the basic cluster. The DR model can, in
principle, lead to quasicrystalline (in particular decagonal and
icosahedral) and to crystalline approximant structures, depending
heavily on the initially chosen seed. The DR growth process can be
interpreted in terms of a cut-and-projection scheme with a peculiar
window function, and the diffraction pattern of DR structures
consists, in general, of Bragg peaks together with a diffuse
background.

The second model in this category, proposed by Janot and Patera
\cite{JanPat} recently, is much simpler than the DR model.  Here,
growth commences from a given seed which determines a ``star'' of
atomic bondings, for instance a decagonal star if the initial cluster
is a decagon with one atom in its center. This vector star then
defines the only possible translations, relative to the surface atoms,
for adding new atoms to the cluster. Growth through the star
short-distance scheme (SSDS) then proceeds as follows. Choosing an
arbitrary surface atom, atoms forming a shifted copy of the basic
initial cluster are added around it. Those of the new atoms that are
too close to already existing sites (the allowed minimum distance is a
parameter of the model) are rejected. Due to the randomness in the
choice of the surface atom, this leads to a family of slightly
different structures. Details on the possible structures, and their
dependence on the choice of the star and the short-distance threshold
are still to be investigated; but it was already shown that also
twinned crystals may result.

\section{Physically Motivated Cluster Growth Models}

While the previously presented models use structureless ``atoms'' to
build up the growing cluster, we now turn to models whose building
blocks already contain information about the symmetry. More
specifically, the orientational models discussed below are random
arrangements of symmetric building blocks (regular decagons or
icosahedra) subject to certain constraints on relative orientations
and distances.

The idea of such a model goes back to Shechtman and Blech
\cite{SheBle}. They tried to explain the observed diffraction patterns
of icosahedral quasicrystals by describing their structure as random
packings of icosahedra, joined along their edges or at their vertices
while preserving orientational symmetry. Albeit their clusters were
rather small (containing about 1000 icosahedra), the computed
diffraction pattern of the vertex-connected random packing gave
reasonable qualitative agreement with experimental results.

This idea was followed up by Stephens and Goldman \cite{SteGold} (see
also the review article by Stephens \cite{Stephens}). Besides vertex-
and edge-connected ``icosahedral glasses'', they also considered
face-connected packings. In particular, they investigated the
size-dependence of the peak widths, performing simulations of up to
$109\, 454$ sites in a sample of cubic shape. The intrinsic peak
widths they obtained turned out to be very small and comparable to
those observed experimentally for icosahedral AlMn. For the majority
of peaks, the cluster size was still insufficient to resolve the
internal width.

Though these results look quite encouraging, Robertson and Moss
\cite{RobMoss} point out that these simple models based on
non-interpenetrating random packings of clusters are unrealistic in
several aspects. First, their density is far too low --- only about
60\% of the bcc packing density.  Another, unrelated problem is the
low, filamentary connectivity of the cluster network. Even reasonably
connected networks were found to be broken up by ``tears'' into
domains.  Accordingly, the phason components behave rather
irregularly, and there is a large increase in the phason fluctuations
with growing patch size. Finally, the peak positions and the peak
shapes calculated from the grown samples do not agree with
experimental powder diffraction patterns of higher-quality
quasicrystals.

Clearly, to improve the situation, one wants to increase connectivity
and suppress the generation of tears. Prior to Robertson and Moss
\cite{RobMoss}, Elser, in a series of papers
\cite{Elser87,Elser88,Elser89}, and Nori et al.\ \cite{NoRoEl} had
already discussed a scenario that strongly improves the quality of the
grown clusters. The main idea, first introduced by Elser for a 2D
random packing model of regular decagons \cite{Elser87}, consists in
introducing a finite temperature into the growth process, thus
allowing rearrangements close to the surface of the growing
sample. More precisely, Elser distinguishes a ``growth region'' with a
non-uniform temperature field, and a ``frozen'' region where no
further rearrangements are allowed. A lattice gas Hamiltonian with
pairwise interactions, including a next-nearest-neighbour term, was
chosen to facilitate the formation of nicely connected networks. In
particular, the allowed cluster-neighbour relationships were taken
from observed approximant phases. The quasicrystal was grown, starting
from a single seed, by moving the growth region with its linear
temperature profile with constant velocity, and applying a Monte Carlo
procedure to the lattice gas Hamiltonian.  The whole procedure easily
translates to the icosahedral case, and the simulations of Elser
\cite{Elser88,Elser89} show that this growth mechanism, with suitable
choices of the parameters, produces well-packed, icosahedrally ordered
structures with intrinsic phason fluctuations that are in good
agreement with the behaviour of peak widths observed in diffraction
experiments. However, due to the anisotropy of the growth conditions,
a small but finite phason strain cannot be avoided.

A somewhat different route was followed by Robertson and Moss
\cite{RobMoss}, imposing additional local constraints on the
neighbourhood of an icosahedron, in this way trying to suppress the
formation of tears and to increase the connectivity. The main idea
consists in allowing a cluster to have only a discrete set of
neighbour distances within its local environment of a certain size. By
suitably tuning their parameters, Robertson and Moss reach packing
fractions and connectivities that are comparable to Elser's, but they
also find a remaining phason strain that does not seem to vanish with
increasing system size.

\begin{figure}
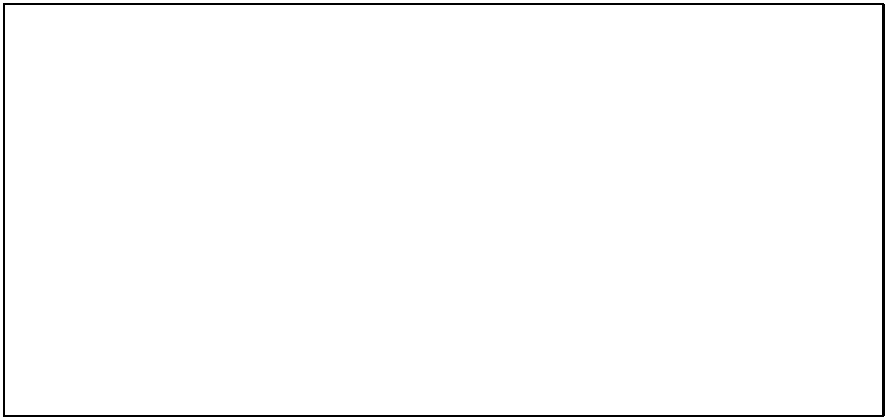

\mpicplace{\textwidth}{55mm}
\caption{Simple equilibrium shapes of icosahedral bond-oriented glasses
(reproduced from \protect\cite{HoJaLiSa}).}
\label{fig7}
\end{figure}

Finally, we would like to point out that facet formation is not a
domain of crystals or ideal quasiperiodic tilings but may also occur
in random tilings and certain classes of bond-oriented glasses (see
\cite{HoJaLiSa,HoLiSaJa,Ho,LeiHen91} for discussions of the possible
\mbox{$T=0$} equilibrium shapes of systems with icosahedral
symmetry). Some simple examples are shown in Fig.~\ref{fig7}.

\section{Random Tiling Growth Models}

Soon after ideal quasiperiodic tilings came into fashion as geometric
models of quasicrystals, Elser \cite{Elser85} pointed out that one may
relax the mathematical constraints of these tilings without losing
their most important features, as for instance their sharp diffraction
peaks. This remark led the way to the so-called random tiling theory
(see \cite{Henley91} for an overview) where tiles are packed randomly
without gaps or overlaps. In a hydrodynamic description, elastic
constants and configurational entropy were introduced. The new degrees
of freedom were (in analogy to modulated crystals) called phasons,
which are long-wavelength modes of the orthogonal coordinates of the
quasicrystal. The difference between the physical properties of these
random tilings compared to those of the ideal quasiperiodic ones --
although mathematically essentially different -- are so small, that
even today it is hard to distinguish between them experimentally, see
\cite{deBo95,Boudard,JoBa,JBR97}.  The local character of the random
tiling scenario made it attractive to growth modelling, having in mind
a cluster decoration of the tiles or the vertices.

Inspired by the structural similarity between the dodecagonal and the
$\sigma$ phase of VNi and VNiSi alloys (which can be described as a
periodic net of squares and equilateral triangles), Kuo et al.\
\cite{KuoFengChen} considered a growth model of dodecagonal
quasicrystals based on the same building blocks, i.e., a
square-triangle random tiling. The ensemble consists of squares and
equilateral triangles only, see \cite{OxHe93}. The growth process
proceeds in attaching complete tiles to a seed following three rules:
(1) only the two tiles considered above are used, (2) squares prefer
triangles on all sides, and (3) when two triangles face each other
they are surrounded by four squares. Whereas rules (1) and (2) --
according to the authors -- yield a modified random tiling with motifs
very similar to those found in high-resolution transmission electron
microscopy (HRTEM) images of dodecagonal phases, the addition of rule
(3) results in a periodic $\sigma$ phase.

In order to obtain a 3D icosahedral random tiling with two linkages
and a high packing density, Henley \cite{Henley91b} developed the
so-called canonical cell tiling (CCT). Its tiles are two irregular
tetrahedra, one irregular square-pyramid and one equilateral prism.
Unfortunately, the CCT is much harder to treat than the primitive
icosahedral tiling \cite{KraNe84}. Motivated by transfer matrix
techniques, Newman et al.\ \cite{NewHenOx95} introduced a method to
produce towers of CCTs with periodic boundary conditions in two
dimensions. The tower construction resembles a growth model. Starting
on a layer of canonical cells which span the entire cross-section, new
vertices of the canonical cells are constructed, without gaps or
overlaps, respecting the boundary conditions. In order to avoid wrong
moves and to save computer time, preference was given to the deepest
lying vertex (w.r.t.\ the surface). For the majority of such growing
surfaces, there is at least one move that is forced by the surrounding
geometry. Following the forced moves, one ends at {\em locally}\/ dead
surfaces where every crevice can be filled at least in two different
ways.  Here, one has to make a choice to proceed. The surface is only
locally dead, i.e., in some cases one ends up with a defective vertex
later on. The catalogue of these dead surfaces is the key feature to
derive estimates for the elastic and entropic properties of the CCT,
see \cite{NewHen95}. The appearance of forced moves and dead surfaces
reminds one of the growth algorithm for the ideal Penrose tiling
described in Sect.~2. However, the canonical cells do not have
matching rules, and their possible arrangements contain a finite
entropy density.

Motivated by the question how close one can approach an {\em
equilibrium}\/ random tiling via a growth process, Joseph and Elser
\cite{JoEl} took a different approach to the square-triangle random
tiling. To every vertex $i$ of a tiling, two variables are associated:
a uniform chemical potential $\mu$ and an angle variable
$\Omega_i$. The chemical potential describes the energy difference of
a bulk vertex relative to the liquid/vapour phase.  The angle
$\Omega_i$ mimics the influence of the surface: it is nothing but the
portion of the full angle around a vertex that is covered by already
completed tiles. Both enter a simple Hamiltonian
\begin{equation}
H = \sum_i (\mu-\Omega_i) \, ,
\label{ham}
\end{equation}
where the sum extends over all vertices of the tiling. In this way,
each bulk vertex has an energy \mbox{$\mu-2\pi$}, whereas surface
vertices have energies \mbox{$\mu-2\pi n/12$} with
$n=0,2,3,4,\ldots,9$ (the case $n=10$ does not occur because the
``missing'' triangle is always complete, and hence the corresponding
vertex belongs to the bulk). Attaching or removing vertices gives rise
to a change in bulk and surface energies. The system will try to
minimize its surface with respect to the bulk. Using (\ref{ham}) as a
penalty function in a Monte Carlo simulation, the authors proposed the
first {\em reversible}\/ growth model with a fluctuating surface,
i.e., vertices may be attached or removed, and detailed balance is
always fulfilled. In the case of the square-triangle tiling, the
geometric constraints were minimal: the minimal distance between
vertices is given by the bond length, and completed tiles may only be
squares or equilateral triangles, while their relative frequency may
vary freely.

Starting from a seed, on the plane or a cylinder, Joseph and Elser
\cite{JoEl} derived a growth/melting boundary of the form
$\mu=2\pi+TS_0$, where $S_0$ is the entropy of the equilibrium
system. The temperature $T$ enters the simulation through the
Boltzmann factor of the Monte Carlo simulation. In this way, it is
possible to measure equilibrium properties of a system via a
non-equilibrium process like growth. For a given temperature
$T<2\pi/3$, the system develops tears for all $\mu<2\pi+TS_0$; and
their separation $\lambda_{\rm tear}$ follows a scaling law
\begin{equation}
\lambda_{\rm tear} \sim (2\pi+TS_0 - \mu)^{-1}
\end{equation}
that is predicted by a phason instability at the surface.
Furthermore, the model shows that the asymmetry in the phason
fluctuations (see Sect.~3) vanishes when approaching the growth
limit. Therefore, in the limit of vanishing growth velocities (i.e.,
$\mu \rightarrow 2\pi+TS_0$), it is possible to produce a grown random
tiling that is, on a given length scale, indistinguishable from a
member of the equilibrium ensemble.  When approaching the growth
limit, the system must exploit all its entropy to grow at all. Thus,
it must align its phason strain closer and closer to zero. In
practice, the system is growing and retreating repeatedly until large,
accidental phason fluctuations are erased.  Using this property, even
the modelling of surface annealing is possible, see \cite{Jo}.

\begin{figure}
\psfig{file=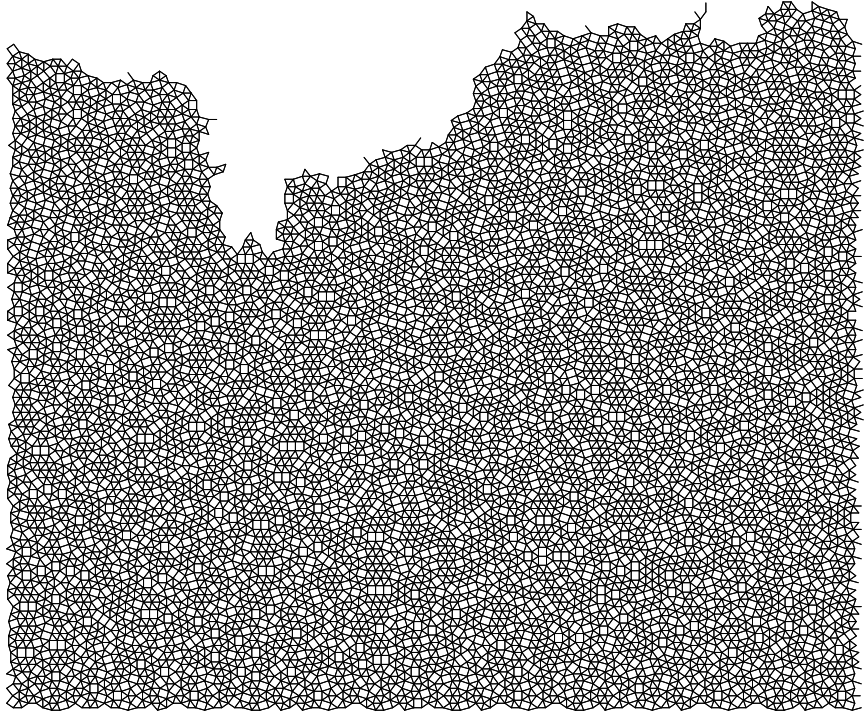,width=\textwidth}
\caption{The unrestricted tenfold rectangle-triangle tiling grown
according to the model of Joseph and Elser \protect\cite{JoEl} at
temperature $T=\pi/2$ with chemical potential $\mu=2\pi+0.160\pi/2$
(rapid growth).  For this case, the growth/melting boundary is
$\mu\approx 2\pi +0.175\pi/2$ \protect\cite{deGier}.}
\label{fig8}
\end{figure}

\begin{figure}
\psfig{file=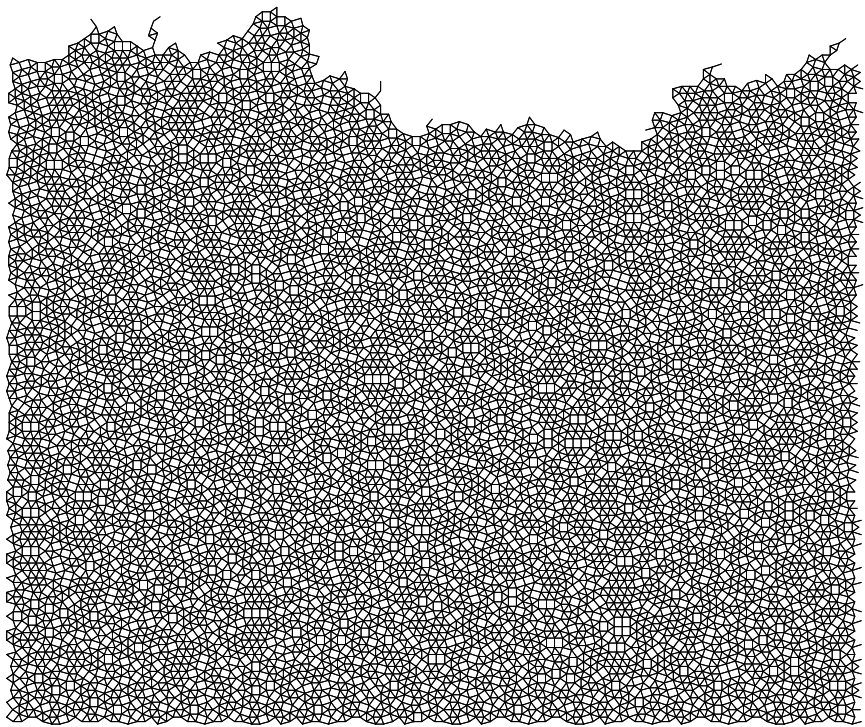,width=\textwidth}
\caption{Same as in Fig.~\protect\ref{fig8}, but this patch was grown
slowly at a chemical potential $\mu=2\pi +0.174\pi/2$.}
\label{fig9}
\end{figure}

The extension to other tilings is straightforward. In Figs.~\ref{fig8}
and \ref{fig9}, we show two cylindrical patches of the unrestricted
tenfold rectangle-triangle random tiling (see \cite{Cockayne}), grown
at different growth velocities. Although this tiling ensemble can be
treated by Bethe-ansatz methods \cite{deGier}, it has not yet been
understood in terms of the conventional random-tiling description, see
\cite{OxMi97}. On the other hand, a restricted (binary) version of
this tiling, which is used to model the decagonal phase of AlPdMn
\cite{OxMi97}, can be treated with random tiling theory without
problems. However, it shows a growth instability due to its binary
restriction which acts as a non-local constraint on the tiling
ensemble \cite{Jo}. Hence growth without leaving the ensemble
appears impossible.

The model of Joseph and Elser \cite{JoEl} shows that, in principle,
there is no need for collective bulk rearrangements after growth and
that uniform phason strains can be made arbitrarily small on
macroscopic length scales.  Usually, bulk rearrangements (in the form
of flips, zippers, \ldots ) are needed to approach equilibrium, but
they can be kinetically very slow.

\section{Concluding Remarks}

To date, it is fair to state that the growth of quasicrystals has not
been understood, and there is still a lot of work to be done in order
to arrive at a physical picture of the growth process that can account
for the experimental observations.  It appears impossible to grow
ideal quasiperiodic structures by purely local algorithms without
incorporating a certain amount of defects. The structure of
quasicrystals may be better described in terms of random tilings (or
other models including a stochastic component), which points at the
importance of entropic contributions in the formation of
quasicrystals.

As we have demonstrated, there are a number of different approaches
towards an understanding of quasicrystal growth. Most models
concentrate on the question of how quasiperiodic order is established,
using a greatly simplified set of degrees of freedom. Clearly, some
remnants of the orientational symmetry are already implicit in all
these models, for instance by allowing only certain atomic positions
or directions of atomic bonding. In this sense, the long-ranged
orientational order in the grown samples is not surprising. What
one really would like to see is a more pronounced order of atomic
positions that can account for the sharp diffraction peaks observed in
experiments. The question whether quasicrystals can be grown by local
deposition of atoms of different species or by attachment of
pre-formed clusters alone remains to be attacked.

Almost all models discussed above employ non-reversible dynamics in
their growth algorithms. Usually, this is said to be motivated by the
rapid quenching processes used to produce quasicrystals, particularly
in the early days (many quasicrystals are still produced by
melt-spinning, with quenching rates of about $10^5\mbox{K}/\mbox{s}$,
but these samples contain many defects; and only after annealing at
about $20$--$50\mbox{K}$ below their melting point, they become
``good'' quasicrystals).  However, the current picture of growth for
crystals suggests diffusion and fluctuation processes on the surface
during growth. There is no reason to believe that quasicrystals are
different in this respect. In our view, this is probably the reason
why most of these models suffer from large anisotropies or phason
fluctuations, creating structures that are far from ideal
quasicrystals or equilibrium random tilings.  Introducing temperature,
and allowing reorganization to occur at least at the surface, the
growing bulk is given the possibility to reconnoiter the
configurational phase space, making full use of its entropy. In this
way, structures that resemble equilibrium random tilings may be
realized by a physical growth process.

Finally, we emphasize that the inflation properties of quasiperiodic
structures allow, in principle, for several physical interpretations
of these models (e.g., ``atoms'' in what we call atomistic models may
be thought of as real atoms, but equally well may be considered as
clusters containing groups of atoms, or even clusters of such
clusters). Thus, also rather simplistic models might well capture the
important ingredients that render possible the formation of
quasicrystals.

\section*{Acknowledgements}

We would like to thank M.\ Baake, V.\ Elser, P.\ Gummelt, C.L.\
Henley, R.\ L\"uck, and J.-B.\ Suck for valuable
comments. Furthermore, we are obliged to V.E.~Dmitrienko and T.L.~Ho
for allowing us to include their figures in this review.  Financial
support from DFG and NSF (Grant No.\ DMR-9412561) is gratefully
acknowledged.

\section*{Note added}

After completion of this review, Dmitrienko et al.\ \cite{DmiAstKle} 
published a modified version of the growth model discussed in 
Sect.~4 \cite{DmiAst}. They replaced the non-local energy 
minimization by a local Monte Carlo step.

\end{document}